\documentclass{ws-procs9x6}

\begin{document}
\newcommand{\be}{\begin{equation}}
\newcommand{\ee}{\end{equation}}
\newcommand{\bea}{\begin{eqnarray}}
\newcommand{\eea}{\end{eqnarray}}
\title{Understanding $D_{sJ}^*(2317)$, $D_{sJ}(2460)$}

\author{F. DE FAZIO}

\address{Istituto Nazionale di Fisica Nucleare, Sezione di Bari, \\
Via Orabona, 4 \\
I-70126 Bari, ITALY\\
E-mail: fulvia.defazio@ba.infn.it}

\maketitle

\abstracts{I briefly review the experimental observations
concerning the charmed mesons  $D_{sJ}^*(2317)$, $D_{sJ}(2460)$
and survey on some of the interpretations proposed in order to
understand their nature. I present an analysis of their decay
modes in the hypothesis that they can be identified with the
scalar and axial vector $s_\ell^P={1 \over 2}^+$ states of $c{\bar
s}$ spectrum ($D_{s0}^*$, $D_{s1}^\prime$). The method is based on
heavy quark symmetries and Vector Meson Dominance ansatz.
Comparison with present data supports the interpretation.}

\section{Introduction}
In April 2003, the BaBar Collaboration  reported the observation
of a narrow peak in the $D_s^+ \pi^0$ invariant mass distribution,
corresponding to a state of mass $2317$ MeV, denoted as
$D_{sJ}^*(2317)$ \cite{Aubert:2003fg}. The state is produced from
charm continuum and the observed width is consistent with the
resolution of the detector, $\Gamma\le 10$ MeV.

A possible quantum number assignment to $D_{sJ}^*(2317)$ is
$J^P=0^+$, as suggested by the angular distribution of the meson
decay with respect to its direction in the $e^+-e^-$ center of
mass frame. This assignment can identify the meson with the scalar
$D_{s0}^*$ state in the spectrum of the $c \bar s$ system.
Considering the masses of the other observed states belonging to
the same system, $D_{s1}(2536)$ and $D_{sJ}(2573)$, the mass of
the scalar $D_{s0}^*$ meson was expected in the range $2.45-2.5$
GeV, therefore $\sim 150$ MeV higher than the observed $2.317$
GeV. A $D_{s0}^*$ meson with such a large mass would be above the
threshold $M_{DK}=2.359$ GeV to strongly decay by $S$-wave Kaon
emission to $DK$, with a consequent broad width. For a  mass below
the $DK$ threshold the meson has to decay by different modes,
namely the isospin-violating $D_s \pi^0$ mode observed by BaBar,
or radiatively. The $J^P=0^+$ assignment excludes the final state
$D_s \gamma$, due to angular momentum and parity conservation;
indeed such a final state has not been observed. On the other
hand, for a scalar $c \bar s$ meson the decay $D_{s0}^* \to D^*_s
\gamma$ is allowed. However, no evidence is reported yet of the
$D_s \gamma \gamma$ final state resulting from the decay chain
$D_{s0}^* \to D^*_s \gamma \to D_s \gamma \gamma$.

Later on, in May 2003, CLEO Collaboration confirmed the BaBar
observation of $D_{sJ}^*(2317)$ with the same features outlined
above; furthermore, CLEO Collaboration observed a second narrow
peak, corresponding to a state with mass 2460 MeV decaying to
$D_s^* \pi^0$ \cite{Besson:2003cp}. Again, the width is compatible
with the detector resolution. Evidence of this second state was
 present in the first  analysis by BaBar Collaboration,
which gave subsequent confirmation of the CLEO
observation\cite{Aubert:2003pe}.  BELLE Collaboration has
confirmed both states\cite{Abe:2003jk}, observing their production
both from charm continuum, both in B decays; more recently also
FOCUS Collaboration\cite{Vaandering:2004ix} has detected of a
narrow peak at 2323 $\pm$ 2 MeV, slightly above the values
obtained by the other three experiments for $D_{sJ}^*(2317)$.

The observation of the decay $D_{sJ}(2460) \to D_s^* \pi^0$
suggests that $D_{sJ}(2460)$ has $J^P=1^+$. This assignment is
supported also by the observation of the mode $D_{sJ}(2460) \to
D_s \gamma$, forbidden to a $0^+$ state, and by the angular
analysis performed by BELLE\cite{Krokovny:2003zq}. Such an
analysis was carried out for $D_{sJ}(2460)$ produced in B decays
and favours the identification of $D_{sJ}(2460)$ with an
axial-vector particle. Production of $D_{sJ}(2460)$ in B decays
was observed also by BaBar\cite{Calderini:2004pd}. However, as in
the case of $D_{sJ}^*(2317)$, the measured mass is below
theoretical expectations for the $1^+$ $c{\bar s}$ state
$D_{s1}^\prime$ and the narrow width contrasts with the expected
broadness of the latter.

These peculiar features of $D_{sJ}^*(2317)$ and $D_{sJ}(2460)$
have prompted a number of  analyses, aimed either at refining
previous results in order to support the $c{\bar s}$
interpretation of $D_{sJ}^*(2317)$ and $D_{sJ}(2460)$, or at
explaining their nature in a different context. The various
interpretations are reviewed in Ref.~\refcite{Colangelo:2004vu}.

Among the non standard scenarios, it has been  often considered
the possibility of a sizeable four-quark component in
$D_{sJ}^*(2317)$ and $D_{sJ}(2460)$. Four-quark states could be
baryonum-like or molecular-like, if they result from bound states
of quarks or of hadrons, respectively, and examples  of the second
kind of states are the often discussed  $f_0(980)$ and $a_0(980)$
when interpreted as $K {\overline K}$ molecules. In the molecular
interpretation, the $D_{sJ}^*(2317)$ could be viewed as a $DK$
molecule\cite{Barnes:2003dj}, an interpretation supported by the
fact that the mass 2.317 GeV is close to the DK threshold, or as a
$D_s \pi$ atom\cite{Szczepaniak:2003vy}. Analogously,
$D_{sJ}(2460)$ would be a $D^*K$ molecule. Mixing between ordinary
$c{\bar s}$ state and a composite state has also been
considered\cite{Browder:2003fk}. No definite answer comes from
lattice QCD, since, according to Ref.~\refcite{Bali:2003jv},
lattice predictions are inconsistent with the simple $q{\bar q}$
interpretation for  $D_{sJ}^*(2317)$, while in
Ref.~\refcite{Dougall:2003hv}  no  exotic scenario is invoked  to
interpret this state.  QCD sum rules are compatible with the
$c{\bar s}$ interpretation\cite{Colangelo:1998ga}.

To understand the structure of a particle one needs  to analyse
its decay modes under definite assumptions and compare the result
with the experimental measurements.  In the following we present
an analysis based on such a strategy  to discuss whether the
identification of $D_{sJ}^*(2317)$ and $D_{sJ}(2460)$ with the two
states $(D_{s0}^*,D_{s1}^\prime)$ is supported by data. To this
end, we compute  the decay modes of a scalar and an axial-vector
particle with masses of $2317$ MeV and 2460 MeV respectively, and
check whether they can be predicted in agreement with the
experimental findings presently available. In particular, the
isospin violating decays to $D_s^{(*)} \pi^0$ should proceed at a
rate larger than the radiative modes, though not exceeding the
experimental upper bounds on the total widths.

\section{Hadronic Modes}
In order to analyze the isospin violating transitions $D_{s0}^*
\to D_s \pi^0$ and $D_{s1}^\prime \to D_s^* \pi^0$, one can use a
formalism that accounts for the heavy quark spin-flavour
symmetries in hadrons containing a single heavy quark, and the
chiral symmetry in the interaction with the octet of light
pseudoscalar states.

In the heavy quark limit, the heavy quark spin $\vec s_Q$ and the
light degrees of freedom total angular momentum $\vec s_\ell$ are
separately conserved. This allows to classify hadrons with a
single heavy quark $Q$ in terms of  $s_\ell$ by collecting them in
doublets the members of which only differ for the relative
orientation of $\vec s_Q$ and $\vec s_\ell$.

The doublets with $J^P=(0^-,1^-)$ and  $J^P=(0^+,1^+)$
(corresponding to $s_\ell^P= {1\over 2}^-$ and $s_\ell^P= {1\over
2}^+$, respectively) can be described by the effective fields
\begin{eqnarray}
H_a &=&
\frac{(1+{\rlap{v}/})}{2}[P_{a\mu}^*\gamma^\mu-P_a\gamma_5]
\label{neg} \\
S_a&=&\frac{1+{\rlap{v}/}}{2}
\left[P_{1a}^{\prime\mu}\gamma_\mu\gamma_5-P_{0a}^*\right]
\label{pos}
\end{eqnarray}
where $v$ is the four-velocity of the meson and $a$ is a light
quark flavour index. In particular in the charm sector the
components of the field  $H_a$  are $P_a^{(*)}=D^{(*)0},D^{(*)+}$
and $D^{(*)}_s$  (for $a=1,2,3$); analogously, the components of
$S_a$ are $P_{0a}^*=D^{*0}_0,D^{*+}_0,D_{s0}^*$ and
$P_{1a}^{\prime}=D^{\prime 0}_1,D^{\prime +}_1,D^\prime_{s1}$.

In terms of these fields it is possible to build up an effective
Lagrange density describing the low energy interactions of heavy
mesons  with the pseudo Goldstone $\pi$, $K$ and $\eta$
bosons\cite{hqet_chir,Casalbuoni:1992gi,Colangelo:1995ph,Casalbuoni:1996pg}:
\begin{eqnarray}
{ L} &=& i\; Tr\{ H_b v^\mu D_{\mu ba} {\overline H}_a \} +
\frac{f_\pi^2}{8} Tr\{\partial^\mu\Sigma\partial_\mu
\Sigma^\dagger \} \nonumber \\ &+&  Tr\{ S_b \;( i \; v^\mu D_{\mu
ba} \; - \; \delta_{ba} \; \Delta)
{\overline S}_a \} \nonumber\\
&+ & \; i \; g \; Tr\{H_b \gamma_\mu \gamma_5 { A}^\mu_{ba}
{\overline H}_a\}
 +  i \; g' \; Tr\{S_b \gamma_\mu \gamma_5 { A}^\mu_{ba} {\overline S}_a\}
\nonumber \\
&+&\,[ i \, h \; Tr\{S_b \gamma_\mu \gamma_5 { A}^\mu_{ba}
{\overline H}_a\}
 \;  + \; h.c.] \;\;\; .  \label{L}
\end{eqnarray}
In  (\ref{L}) $\overline H_a$ and  $\overline S_a$ are defined as
$\overline H_a= \gamma^0 H^\dagger_a \gamma^0$ and $\overline S_a=
\gamma^0 S^\dagger_a \gamma^0$;   all the heavy field operators
contain a factor $\sqrt{M_P}$ and have dimension $3/2$. The
parameter $\Delta$ represents the mass splitting between positive
and negative parity states.

The $\pi$, $K$ and $\eta$ pseudo Goldstone bosons are included in
the effective lagrangian (\ref{L}) through the field
$\displaystyle \xi=e^{i { M} \over f}$ that represents a unitary
matrix describing the pseudoscalar octet, with
\begin{equation}
{ M}= \left (\begin{array}{ccc}
\sqrt{\frac{1}{2}}\pi^0+\sqrt{\frac{1}{6}}\eta & \pi^+ & K^+\nonumber\\
\pi^- & -\sqrt{\frac{1}{2}}\pi^0+\sqrt{\frac{1}{6}}\eta & K^0\\
K^- & {\bar K}^0 &-\sqrt{\frac{2}{3}}\eta
\end{array}\right ) \label{M}
\end{equation}
and $f\simeq f_{\pi}$. In eq.(\ref{L}) $\Sigma=\xi^2$, while the
operators $D$ and $ A$  are given by:
\begin{eqnarray}
D_{\mu ba}&=&\delta_{ba}\partial_\mu+{ V}_{\mu ba}
=\delta_{ba}\partial_\mu+\frac{1}{2}\left(\xi^\dagger\partial_\mu
\xi
+\xi\partial_\mu \xi^\dagger\right)_{ba}\\
{ A}_{\mu ba}&=&\frac{1}{2}\left(\xi^\dagger\partial_\mu \xi-\xi
\partial_\mu \xi^\dagger\right)_{ba} \; .
\end{eqnarray}

The strong interactions between the heavy $H_a$ and $S_a$ mesons
with the light pseudoscalar mesons are thus governed, in the heavy
quark limit,  by three dimensionless couplings: $g$, $h$ and
$g^\prime$. In particular, $h$ describes the coupling between a
member of the $H_a$ doublet and one of the $S_a$ doublet to a
light pseudoscalar meson, and is the one relevant for our
discussion.

Isospin violation enters in the low energy Lagrangian of $\pi$, K
and $\eta$ mesons through the mass term \be { L}_{mass} = {\tilde
\mu f^2 \over 4} Tr\{\xi m_q \xi +\xi^\dagger m_q \xi^\dagger\}
\label{Lmass} \ee with $m_q$ the light quark mass matrix:
\begin{equation}
m_q= \left (\begin{array}{ccc} m_u&0&0\nonumber\\0&m_d&0\\ 0&0&m_s
\end{array}\right ) \,\,\, .
\end{equation}
In addition to the light meson mass terms, $ L_{mass}$ contains an
interaction term between $\pi^0$ ($I=1$) and  $\eta$ ($I=0$)
mesons: ${ L}_{mixing}={\tilde \mu \over 2} {m_d-m_u\over \sqrt 3}
\pi^0 \eta$  which vanishes in the limit $m_u=m_d$. Let us focus
on the mode $D_{s0}^* \to D_s \pi^0$. As in the case of $D_s^* \to
D_s \pi^0$ studied in Ref.~\refcite{Cho:1994zu}, the isospin
mixing term can drive such a  transition\footnote{Electromagnetic
contributions to $D_{s0}^* \to D_s \pi^0$ are expected to be
suppressed with respect to the strong interaction mechanism
considered here.}. The amplitude $A(D_{s0}^*\to D_s \pi^0)$ is
simply written in terms of $A(D_{s0}^*\to D_s \eta)$ obtained from
(\ref{L}), $A(\eta\to \pi^0)$ from (\ref{Lmass}) and the $\eta$
propagator that puts the strange quark mass in the game. The
resulting expression for the decay amplitude involves the coupling
$h$ and the suppression factor $(m_d-m_u)/(m_s-{m_d+m_u \over 2})$
accounting for isospin violation, so that the width
$\Gamma(D_{s0}^* \to D_s \pi^0)$ reads: \be \Gamma(D_{s0}^* \to
D_s \pi^0) = {1 \over 16 \pi} {h^2 \over f^2} {M_{D_s} \over
M_{D_{s0}^*}} \Big( {m_d-m_u \over m_s-{m_d+m_u \over 2}}\Big)^2
(1 + {m_{\pi^0}^2 \over |\vec p_{\pi^0}|^2}) |\vec p_{\pi^0}|^3
\,\,\, .\label{gamma1} \ee As for $h$, the result of  QCD sum rule
analyses of various heavy-light quark current correlators is $|h|
= 0.6 \pm 0.2$ \cite{Colangelo:1995ph}. Using the central value,
together with the factor $(m_d-m_u)/(m_s-{m_d+m_u \over 2}) \simeq
{1 \over 43.7}$ \cite{Gasser:1984gg} and $f=f_\pi=132$ MeV we
obtain\cite{Colangelo:2003vg}: \be \Gamma(D_{s0}^* \to D_s
\pi^0)=7 \pm 1 \,\, KeV \label{had-ris} \,. \ee

The analogous calculation for $D_{s1}^\prime \to D_s^* \pi^0$
provides  the result\cite{Colangelo:2004vu}: \be
\Gamma(D_{s1}^\prime \to D_s^* \pi^0)= 7 \pm 1\,\, KeV
\label{had-ris1} \,. \ee

\section{Radiative Modes}
Let us now turn to the calculation of radiative decay rates. We
describe the procedure considering the mode $D_{s0}^* \to D_s^*
\gamma$, the amplitude of which has the form:
  \be { A}(D_{s0}^* \to D^*_s \gamma)= e \,
d \, [(\epsilon^* \cdot \eta^*)(p \cdot k)-(\eta^* \cdot p)
(\epsilon^* \cdot k)] \label{amptot} \,, \ee where $p$ is the
$D_{s0}^*$ momentum, $\epsilon$ the $D^*_s$ polarization vector,
and $k$ and $\eta$ the photon momentum and polarization. The
corresponding decay rate is: \be \Gamma(D_{s0}^* \to D_s^*
\gamma)=\alpha |d|^2 |\vec{k}|^3 \,\,\, . \label{radwid} \ee The
parameter $d$ gets contributions from the photon couplings to the
light quark part $e_s {\bar s}\gamma_\mu s$ and to the heavy quark
part $e_c {\bar c}\gamma_\mu c$ of the electromagnetic current,
$e_s$ and $e_c$ being strange and charm quark charges in units of
$e$. Its general structure is: \be d=d^{(h)}+d^{(\ell)}={e_c \over
\Lambda_c} + {e_s \over \Lambda_s} \, , \label{mu} \ee where
$\Lambda_a$ ($a=c,s$) have dimension of a mass. Such a structure
is already known from the constituent quark model. In the case of
$M1$ heavy meson transitions, an analogous structure predicts a
relative suppression of the radiative rate of the charged $D^*$
mesons  with respect to the neutral
one\cite{Eichten:1979ms,Amundson:1992yp,Colangelo:1994jc,Colangelo:1993zq},
suppression that has  been experimentally
confirmed\cite{Hagiwara:fs}. From  (\ref{radwid},\ref{mu}) one
could expect a small width for the transition $D_{s0}^* \to D_s^*
\gamma$, to be compared to the hadronic width $D_{s0}^* \to D_s
\pi^0$ which is suppressed as well.

In order to determine the amplitude of  $D_{s0}^* \to D_s^*
\gamma$ we follow a method based again on the use of heavy quark
symmetries, together with the vector meson dominance (VMD)
ansatz\cite{Amundson:1992yp,Colangelo:1993zq}. We first consider
the coupling of the photon to the heavy quark part of the e.m.
current. The matrix element $ \langle D^*_s(v^\prime,
\epsilon)|{\bar c}\gamma_\mu c|D_{s0}^*(v) \rangle $ ($v$,
$v^\prime$ meson four-velocities) can be computed in the heavy
quark limit, matching the QCD ${\bar c}\gamma_\mu c$ current onto
the corresponding HQET expression\cite{Falk:1990pz}: \be
J_\mu^{HQET}=\bar h_v [ v_\mu + {i \over 2 m_Q}( \overrightarrow
\partial_\mu - \overleftarrow \partial_\mu) + {i \over 2 m_Q}
\sigma_{\mu\nu} ( \overrightarrow \partial^\nu + \overleftarrow
\partial^\nu) + \dots]h_v \ee where $h_v$ is the effective field
of the heavy quark. For transitions involving $D_{s0}^*$ and
$D_s^*$, and for $v=v^\prime$ ($v \cdot v^\prime=1$), the matrix
element of $J_\mu^{HQET}$ vanishes. The consequence is that
$d^{(h)}$ is proportional to the inverse heavy quark mass $m_Q$
and presents a suppression factor since in the physical case
$v\cdot v^\prime=(m_{D_{s0}^*}^2+m_{D^*_s}^2)/2
m_{D_{s0}^*}m_{D^*_s}=1.004$. Therefore, we neglect $d^{(h)}$ in
(\ref{mu}).

To evaluate the coupling of the photon to the light quark part of
the electromagnetic current we invoke the VMD ansatz and consider
the contribution of the intermediate $\phi(1020)$: \bea && \langle
D^*_s(v^\prime, \epsilon)|{\bar s}\gamma_\mu s|D_{s0}^*(v)
\rangle= \label{vmd} \\ &&
 \sum_{\lambda}
 \langle D^*_s(v^\prime,
\epsilon) \phi(k, \epsilon_1(\lambda))|D_{s0}^*(v) \rangle {i
\over k^2- M_\phi^2} \langle 0 |{\bar s}\gamma_\mu s|\phi(k,
\epsilon_1(\lambda)) \rangle \nonumber  \eea with $k^2=0$ and
$\langle 0 |{\bar s}\gamma_\mu s|\phi(k, \epsilon_1)
\rangle=M_\phi f_\phi \epsilon_{1 \mu}$. The experimental value of
$f_\phi$ is $f_\phi=234 $ MeV. The matrix element $\langle
D^*_s(v^\prime, \epsilon) \phi(k, \epsilon_1)|D_{s0}^*(v) \rangle$
describes the strong interaction of a light vector meson ($\phi$)
with two heavy mesons ($D^*_s$ and $D_{s0}^*$). It  can also be
obtained through a low energy  lagrangian in which the heavy
fields $H_a$ and $S_a$ are coupled, this time, to the octet of
light vector mesons\footnote{The standard $\omega_8-\omega_0$
mixing is assumed, resulting in a pure $\bar s s$ structure for
$\phi$.}. The Lagrange density is set up using the hidden gauge
symmetry method\cite{Casalbuoni:1992gi}, with the light vector
mesons collected in a 3 $\times$ 3  matrix ${\hat \rho}_\mu$
analogous to ${ M}$ in (\ref{M}). The lagrangian\footnote{The role
of other possible structures in the effective lagrangian
contributing to radiative decays is discussed in
Ref.~\refcite{Colangelo:1993zq}.} reads
as\cite{Casalbuoni:1992dx}: \be { L}^\prime= i \, \hat \mu \, Tr
\{ {\bar S}_a H_b \sigma^{\lambda \nu} V_{\lambda \nu}(\rho)_{ba}
\} +h.c. \label{vec-lag} \,, \ee with $V_{\lambda
\nu}(\rho)=\partial_\lambda \rho_\nu-\partial_\nu
\rho_\lambda+[\rho_\lambda,\rho_\nu]$ and  $\rho_\lambda=i{g_V
\over \sqrt{2}}{\hat \rho}_\lambda$, $g_V$ being fixed  to
$g_V=5.8$ by the KSRF rule\cite{Ksrf}. The coupling $\hat \mu$ in
(\ref{vec-lag}) is constrained to $\hat \mu=-0.1  \, GeV^{-1}$ by
the analysis of the $D \to K^*$ semileptonic transitions induced
by the axial weak
current\cite{Casalbuoni:1992dx,Casalbuoni:1996pg}. The resulting
expression for ${1 \over {\Lambda}_s}$ is:
\be {1 \over {\Lambda}_s}=-4\hat\mu{g_V \over
 \sqrt{2}} \sqrt{{M_{D^*_s}\over M_{D_{s0}^*}}}{f_\phi \over M_\phi}
\,\,\, .\label{mag-mom} \ee

The numerical result for the radiative
width\cite{Colangelo:2003vg}: \be \Gamma(D_{s0}^* \to D_s^*
\gamma) =0.85 \pm 0.05 \,\, KeV \label{rad-ris} \,  \ee shows that
the hadronic $D_{s0}^* \to D_s^* \pi^0$ transition is  more
probable than the radiative mode $D_{s0}^* \to D_s^* \gamma$, at
odds with the case of the $D_s^*$ meson, where the radiative mode
dominates the decay rate. In particular, if we assume that the two
modes essentially saturate the  $D_{s0}^*$ width, we have
$\Gamma(D_{s0}^*)=8 \pm 1$ KeV.
  As for the two radiative modes allowed
for $D_{sJ}(2460)$, one finds\cite{Colangelo:2004vu}:
 \bea  \Gamma(D_{s1}^\prime \to D_s
\gamma)= 3.3 \pm 0.6  \,\,\, KeV &\hskip 0.4 cm&
\Gamma(D_{s1}^\prime \to D_s^* \gamma)= 1.5 \,\,\, KeV \eea
\noindent which in turn give a total width $\Gamma(D_{s1}^\prime)=
12 \pm 1$ KeV.

\section{Comparison with other approaches}

The results of the previous two sections show that, within the
described approach, the observed hierarchy of hadronic  versus
radiative modes is realized, supporting the identification of
$D_{sJ}^*(2317)$ and $D_{sJ}(2460)$ with
$(D_{s0}^*,D_{s1}^\prime)$. Other analyses have followed the same
strategy of computing decay rates of the two narrow states in
order to understand their structure. In Table \ref{width} we
compare our results with the outcome of other approaches based on
the $c \bar s$ picture as well. Analyses in which the states are
assumed to have an exotic structure provide  larger values for the
widths (${\it O}(10^2)$ KeV)\cite{Cheng:2003kg}.

\begin{table}[h]
\tbl{Estimated width (KeV) of $D_{s0}^*$ and $D_{s1}^\prime$,
using the $c \bar s$ picture.
 The results in column
[35] are obtained using experimental inputs from
 Belle (Focus).} {\footnotesize
\begin{tabular}{  l  c c  c  c c } \hline
Decay mode & [\refcite{eichten}] & [\refcite{Godfrey:2003kg}] &
[\refcite{Colangelo:2003vg,Colangelo:2004vu}]& [\refcite{:2003dp}]
& [\refcite{Azimov:2004xk}]
\\[1ex] \hline $D_{s0}^* \to D_s \pi^0$ & 21.5  & $\simeq 10$  & $7 \pm 1$
 & 16  & \begin{tabular}{c}  $129 \pm 43 $ \\ $(109\pm 16)$
\end{tabular}      \\[1ex]
$D_{s0}^* \to D_s^* \gamma$ & 1.74  & 1.9  & $0.85 \pm 0.05$
 & 0.2  & $\le 1.4$  \\[1ex]
\hline $D_{s1}^\prime \to D_s^* \pi^0$ & 21.5  & $\simeq 10$ & $7
\pm 1$ & 32  & \begin{tabular}{c}  $187 \pm 73$ \\ $(7.4\pm 2.3)$
\end{tabular}   \\[1ex]
$D_{s1}^\prime \to D_s \gamma$ & 5.08  & 6.2& $3.3 \pm 0.6$ &
& $\le 5$   \\[1ex]
$D_{s1}^\prime \to D_s^* \gamma$ & 4.66  & 5.5& 1.5  & &   \\[1ex]
\hline
\end{tabular}\label{width} }
\vspace*{-13pt}
\end{table}

In particular, we observe that conclusions analogous to those
presented above have been reached in  Ref.~\refcite{eichten},
which is based on the observation  that heavy-light systems should
appear as parity-doubled, i.e. in pairs differing for parity and
transforming according to a linear representation of chiral
symmetry. In particular, the doublet composed by the states having
$J^P=(0^+,1^+)$ can be considered as the chiral partner of that
with $J^P=(0^-,1^-)$ \footnote{This idea was first suggested in
Ref.~\refcite{Nowak:1992um} in order to obtain a consistent
implementation of chiral symmetry  and reconsidered also in
Ref.~\refcite{Nowak:2003ra}.}. Since our calculation is based on a
different method, the $s_\ell^P=
 {1\over 2}^-$ and $s_\ell^P= {1 \over 2}^+$ doublets being treated
as uncorrelated multiplets, we find the agreement noticeable.

\section{Conclusions and perspectives}

We presented the calculation of hadronic and radiative decay rates
of $D_{sJ}^*(2317)$ and $D_{sJ}(2460)$ in a framework based on
heavy quark symmetries and on the Vector Meson Dominance ansatz.
This analysis shows that  the observed narrow widths and the
enhancement of the $D_s^{(*)} \pi^0$ decay modes are compatible
with the identification of $D_{sJ}^*(2317)$ and $D_{sJ}(2460)$
with the states belonging to the $J^P_{s_\ell}=(0^+,1^+)_{1 \over
2}$ doublet of the $c \bar s$ spectrum. Nevertheless, unanswered
questions remain, such as the near equality of the masses of
$D_{sJ}^*(2317)$ and $D_{sJ}(2460)$ with their  non-strange
partners. The missing  evidence of  the radiative mode
$D^*_{sJ}(2317) \to D_s^* \gamma$ is another puzzling aspect
deserving further experimental investigations.

The quantum number assignment has  a rather straightforward
consequence concerning the doublet of scalar and axial vector
mesons in the $b \bar s$ spectrum. Since the mass splitting
between $B$ and $D$ states is similar to the corresponding mass
splitting between $B_s$ and $D_s$ states, such mesons should be
below the $BK$ and $B^*K$ thresholds, thus producing narrow peaks
in $B_s \pi^0$ and $B_s^* \pi^0$ mass distributions\footnote{A
collection of  predictions for the masses of $s_\ell^P=1/2^+$
$b{\bar s}$ mesons can be found in Ref.
\refcite{Colangelo:2004vu}; see also Ref.
\refcite{vanBeveren:2004bz}.}.

In conclusion, we expect that new experimental results will
continue to enrich this scenario. Actually, another excited
$c{\bar s}$ meson has been recently observed by Selex
Collaboration\cite{Evdokimov:2004iy}, motivating further studies
in this exciting period for hadron spectroscopy.

\section*{Acknowledgments}
I warmly thank the Organizers for inviting me to present this talk
and for their very kind hospitality. I also thank P. Colangelo and
R. Ferrandes for collaboration on the topics discussed above.
 Partial support from the EC Contract No.
HPRN-CT-2002-00311 (EURIDICE) is acknowledged.


\begin{thebibliography}{0}
\bibitem{Aubert:2003fg}
B.~Aubert {\it et al.}  [BABAR Collaboration], {\it Phys. Rev.
Lett.} {\bf 90}, 242001 (2003).

\bibitem{Besson:2003cp}
D.~Besson {\it et al.}  [CLEO Collaboration], {\it Phys. Rev.}
{\bf D68}, 032002 (2003).

\bibitem{Aubert:2003pe}
B.~Aubert {\it et al.}  [BABAR Collaboration], {\it Phys. Rev.}
{\bf D69}, 031101 (2004).

\bibitem{Abe:2003jk}
K.~Abe {\it et al.}, {\it Phys. Rev. Lett.}  {\bf 92}, 012002
(2004).

\bibitem{Vaandering:2004ix}
E.~Vaandering, arXiv:hep-ex/0406044.


\bibitem{Krokovny:2003zq}
P.~Krokovny {\it et al.}  [Belle Collaboration], {\it Phys. Rev.
Lett.}  {\bf 91}, 262002 (2003).
 P.~Krokovny, arXiv:hep-ex/0310053.

\bibitem{Calderini:2004pd}
G.~Calderini  [BaBar Collaboration], arXiv:hep-ex/0405081.

\bibitem{Colangelo:2004vu}
P.~Colangelo, F.~De Fazio and R.~Ferrandes,
arXiv:hep-ph/0407137.


\bibitem{Barnes:2003dj}
T.~Barnes, F.~E.~Close and H.~J.~Lipkin, {\it Phys. Rev.}  {\bf
D68}, 054006 (2003);
H.~J.~Lipkin, {\it Phys. Lett.}  {\bf B580}, 50 (2004);
P.~Bicudo, arXiv:hep-ph/0401106.

\bibitem{Szczepaniak:2003vy}
A.~P.~Szczepaniak, {\it Phys. Lett.}  {\bf B567}, 23 (2003).


\bibitem{Browder:2003fk}
T.~E.~Browder, S.~Pakvasa and A.~A.~Petrov, {\it Phys. Lett.} {\bf
B578}, 365 (2004);
S.~Nussinov, arXiv:hep-ph/0306187.

\bibitem{Bali:2003jv}
G.~S.~Bali, {\it Phys. Rev.}  {\bf D68}, 07150 (2003).

\bibitem{Dougall:2003hv}
A.~Dougall, R.~D.~Kenway, C.~M.~Maynard and C.~McNeile  [UKQCD
                  Collaboration],
{\it Phys. Lett.}  {\bf B569}, 41 (2003).

\bibitem{Colangelo:1998ga}
P.~Colangelo, F.~De Fazio and N.~Paver,  {\it Phys. Rev.} {\bf
D58}, 116005 (1998);
Y.~B.~Dai, C.~S.~Huang, C.~Liu and S.~L.~Zhu,
 {\it Phys. Rev.}  {\bf D68}, 114011 (2003).

\bibitem{hqet_chir}
M.~B. Wise, {\it Phys. Rev. } {\bf D45}, R2188 (1992); G.Burdman
and J.~F.  Donoghue, {\it Phys. Lett.}  {\bf  B280}, 287 (1992);
P. Cho, {\it Phys. Lett.} {\bf  B285}, 145 (1992); H.-Y.Cheng,
C.-Y. Cheung, G.-L. Lin, Y.~C. Lin and H.-L. Yu, {\it Phys. Rev.}
{\bf D46}, 1148 (1992).

\bibitem{Casalbuoni:1992gi}
R.~Casalbuoni, A.~Deandrea, N.~Di Bartolomeo, R.~Gatto,
F.~Feruglio and G.~Nardulli, {\it Phys.\ Lett.} {\bf B292}, 371
(1992).

\bibitem{Colangelo:1995ph}
P.~Colangelo, F.~De Fazio, G.~Nardulli, N.~Di Bartolomeo and
R.~Gatto, {\it Phys.\ Rev.} {\bf D52}, 6422 (1995);
P.~Colangelo and F.~De Fazio, {\it Eur. Phys. J.}  {\bf C4}, 503
(1998).

\bibitem{Casalbuoni:1996pg}
For a review see: R.~Casalbuoni, A.~Deandrea, N.~Di Bartolomeo,
R.~Gatto, F.~Feruglio and G.~Nardulli, {\it Phys. Rept.}  {\bf
281}, 145 (1997).


\bibitem{Cho:1994zu}
P.~L.~Cho and M.~B.~Wise, {\it Phys. Rev.} {\bf D49}, 6228 (1994).


\bibitem{Gasser:1984gg}
J.~Gasser and H.~Leutwyler, {\it Nucl. Phys.}  {\bf B250}, 465
(1985).

\bibitem{Colangelo:2003vg}
P.~Colangelo and F.~De Fazio, {\it Phys. Lett.} {\bf B570}, 180
(2003).


\bibitem{Eichten:1979ms}
E.~Eichten, K.~Gottfried, T.~Kinoshita, K.~D.~Lane and T.~M.~Yan,
{\it Phys. Rev.}  {\bf D21}, 203 (1980).


\bibitem{Amundson:1992yp}
J.~F.~Amundson {\it et al.}, {\it Phys. Lett.} {\bf B296}, 415
(1992);
P.~L.~Cho and H.~Georgi, {\it Phys. Lett.} {\bf B296},408 (1992)
[Erratum-ibid.\  {\bf B300}, 410 (1993)].

\bibitem{Colangelo:1994jc}
P.~Colangelo, F.~De Fazio and G.~Nardulli, {\it Phys. Lett.} {\bf
B334}, 175 (1994).

\bibitem{Colangelo:1993zq}
P.~Colangelo, F.~De Fazio and G.~Nardulli, {\it Phys. Lett.} {\bf
B316}, 555 (1993).


\bibitem{Hagiwara:fs}
K.~Hagiwara {\it et al.}  [Particle Data Group Collaboration],
{\it Phys. Rev.}  {\bf D66}, 010001 (2002).


\bibitem{Falk:1990pz}
A.~F.~Falk, B.~Grinstein and M.~E.~Luke,  {\it Nucl. Phys.}  {\bf
B357}, 185 (1991).

\bibitem{Casalbuoni:1992dx}
R.~Casalbuoni, A.~Deandrea, N.~Di Bartolomeo, R.~Gatto,
F.~Feruglio and G.~Nardulli, {\it Phys. Lett.}  {\bf B299}, 139
(1993).


\bibitem{Ksrf}
K.~Kawarabayashi and M.~Suzuki, {\it Phys. Rev. Lett.}  {\bf 16},
255 (1966);
Riazuddin and Fayazuddin, {\it Phys. Rev.} {\bf 147}, 1071 (1966).

\bibitem{Cheng:2003kg}
H.~Y.~Cheng and W.~S.~Hou, {\it Phys. Lett.}  {\bf B566}, 193
(2003);
S.~Ishida, M.~Ishida, T.~Komada, T.~Maeda, M.~Oda, K.~Yamada and
I.~Yamauchi, arXiv:hep-ph/0310061.

\bibitem{eichten}
W.~A.~Bardeen, E.~J.~Eichten and C.~T.~Hill, {\it Phys. Rev.} {\bf
D68}, 054024 (2003).

\bibitem{Godfrey:2003kg}
S.~Godfrey, {\it Phys. Lett.}  {\bf B568}, 254 (2003).

\bibitem{:2003dp}
Fayyazuddin and Riazuddin, arXiv:hep-ph/0309283.


\bibitem{Azimov:2004xk}
Y.~I.~Azimov and K.~Goeke, arXiv:hep-ph/0403082.



\bibitem{Nowak:1992um}
M.~A.~Nowak, M.~Rho and I.~Zahed, {\it Phys. Rev.}  {\bf D48},
4370 (1993);
W.~A.~Bardeen and C.~T.~Hill, {\it Phys. Rev.}  {\bf D49}, 409
(1994).

\bibitem{Nowak:2003ra}
M.~A.~Nowak, M.~Rho and I.~Zahed, arXiv:hep-ph/0307102.

\bibitem{vanBeveren:2004bz}
E.~van Beveren and G.~Rupp,
arXiv:hep-ph/0406242.

\bibitem{Evdokimov:2004iy}
A.~V.~Evdokimov  [SELEX Collaboration],
arXiv:hep-ex/0406045.



\end{thebibliography}
\end{document}